\documentclass[aps,noshowkeys]{revtex4}
\usepackage{epsfig}
\usepackage{graphicx}
\usepackage{bm}
\usepackage{amssymb,amsmath,euscript,esint,mathtools}

\begin{document}
\newcommand{\Br}[1]{(\ref{#1})}
\newcommand{\Eq}[1]{Eq.\ (\ref{#1})}
\newcommand{\frc}[2]{\raisebox{1pt}{$#1$}/\raisebox{-1pt}{$#2$}}
\newcommand{\frcc}[2]{\raisebox{0.3pt}{$#1$}/\raisebox{-0.3pt}{$#2$}}
\newcommand{\frccc}[2]{\raisebox{1pt}{$#1$}\big/\raisebox{-1pt}{$#2$}}
\title{Accurate Evaluation of $\bm{\mathcal{P}}$,$\bm{\mathcal{T}}$-odd Faraday Effect in Atoms of Xe and Hg}
\author{D. V. Chubukov$^{1,2}$, L. V. Skripnikov$^{2,1}$, L. N. Labzowsky$^{1,2}$, V. N. Kutuzov$^1$ and S. D. Chekhovskoi$^1$}
\affiliation{$^1$ Department of Physics, St. Petersburg State University,
7/9 Universitetskaya Naberezhnaya, St. Petersburg 199034, Russia \\
$^2$Petersburg Nuclear Physics Institute named by B.P. Konstantinov of
National Research Centre ``Kurchatov Institut'', St. Petersburg, Gatchina 188300, Russia \\
}

\begin{abstract}
Accurate evaluation of the $\mathcal{P}$,$\mathcal{T}$-odd Faraday effect (rotation of the polarization plane for the light propagating through a medium in presence of an external electric field) is presented. This effect can arise only due to the $\mathcal{P}$,$\mathcal{T}$-odd ($\mathcal{P}$ - space parity, $\mathcal{T}$ - time reflection) interactions and is different from the ordinary Faraday effect, i.e. the light polarization plane rotation in an external magnetic field. The rotation angle is evaluated for the ICAS (intracavity absorption spectroscopy) type experiments with Xe and Hg atoms. The results show that Hg atom may become good candidate for a search for the $\mathcal{P}$,$\mathcal{T}$-odd effects in atomic physics.
\end{abstract}
\maketitle
\section{Introduction}

A search for the $\mathcal{P}$,$\mathcal{T}$-odd effects in the low energy physics started with the paper \cite{Pur50} where the possibility to observe the electric dipole moment (EDM) of the neutron was first discussed. The existence of the EDM for any particle or closed system of particles violates the space parity ($\mathcal{P}$) and time invariance ($\mathcal{T}$) conservation. Later an another $\mathcal{P}$,$\mathcal{T}$-odd effect was described: $\mathcal{P}$,$\mathcal{T}$-odd interaction of electron and nucleus in atomic systems \cite{San75}, \cite{Gor79}. Both effects can be observed in an external electric field and cannot be distinguished from each other in any particular experiment with any atom or molecule. However they can be distinguished in a series of experiments with different species. References to the numerous papers on the subject can be found in the book \cite{Khrip91} and the review \cite{Gin04}.

At the moment the experimental limitations for the particles EDMs are most advanced for the electrons since the electron EDM ($e$EDM) is greatly enhanced in heavy atoms and especially in heavy diatomic molecules. This is true also for the $\mathcal{P}$,$\mathcal{T}$-odd electron-nucleus interaction which is convenient to express via the equivalent $e$EDM. An equivalent $e$EDM in any atomic system can be defined as the $e$EDM that leads to the same linear Stark  shift in the same external electric field as the given electron-nucleus $\mathcal{P}$,$\mathcal{T}$-odd interaction. The most restrictive bounds for the $e$EDM were established in the experiments with Tl atom ($d_e<1.6\times 10^{-27}$  $e$ cm) \cite{Reg02}, YbF molecule ($d_e<1.05\times 10^{-27}$ $e$ cm) \cite{Hud11}, ThO molecule ($d_e< 0.87 \times 10^{-28}$ $e$ cm \cite{ACME13}, $d_e<1.1\times 10^{-29}$ $e$ cm \cite{ACME18}), and HfF$^+$ molecular ion ($d_e<1.3\times 10^{-28}$ $e$ cm) \cite{Cair17}. Here $e$ is the electron charge. For the extraction of $d_e$ values from the experimental data the theoretical calculations of the enhancement coefficients are required. These calculations were performed for Tl in \cite{Liu92,Dzuba09,Por12,Chub18}, for YbF in \cite{Quiney:98, Parpia:98, Mosyagin:98}, for ThO in \cite{Skripnikov:13c,Skripnikov:15a,Skripnikov:16b,Fleig:16}  and for HfF$^+$ in \cite{Petrov:07a,Skripnikov:17c, Fleig:17, Petrov:18b}.       

The theoretical prediction of the $e$EDM value is rather uncertain. Within the Standard Model (SM) none of these predictions promises for the $d_e$ magnitude the value larger than $10^{-38}$ $e$ cm \cite{Pos91} (i.e. 9 orders of magnitude smaller than the recent experimental bound). We do not discuss here the possible consequences of ``new physics''. The largest prediction for effective $d_e^{\text{eff}}$ originating from the $\mathcal{P}$,$\mathcal{T}$-odd two-photon exchange between an electron and a nucleus in atomic systems was estimated in \cite{Pos14} as $10^{-38}$ $e$ cm. In the same paper an $e$EDM $d_e$ was estimated to be much smaller than the value for $d_e^{\text{eff}}$. Another model for the $\mathcal{P}$,$\mathcal{T}$-odd electron-nucleus interaction in atomic systems via exchange by the Higgs boson was discussed in \cite{Chub16}. The predictions for the $d_e^{\text{eff}}$ within this model are also rather uncertain.

In the modern experiments on the search for the $\mathcal{P}$,$\mathcal{T}$-odd effects in atomic and molecular systems either the shift of the magnetic resonance in an electric field \cite{Reg02} or the electron spin precession in an external electric field \cite{Hud11,ACME13,ACME18,Cair17} had to be observed. Due to the very large gap between the minimum experimental bound and the maximum theoretical prediction within the SM the other possible methods of observation of the $\mathcal{P}$,$\mathcal{T}$-odd effects in atomic and molecular systems may be of interest. One of such methods is the rotation of the polarization plane of the light propagating through a medium in the presence of an external electric field. This method can be called the $\mathcal{P}$,$\mathcal{T}$-odd Faraday effect. An existence of such an effect was first mentioned in \cite{Baran78} and the possibility to observe it was studied theoretically and experimentally (see the short review on the subject in \cite{Bud02}). Recently, a possible observation of the $\mathcal{P}$,$\mathcal{T}$-odd Faraday effect by the methods of intracavity absorption spectroscopy (ICAS) \cite{Boug14,Baev99,Dur10} was discussed in \cite{Chub17}. The ICAS experiments are most suitable for the observation of the $\mathcal{P}$,$\mathcal{T}$-odd Faraday effect. In particular, in \cite{Boug14} an experiment on the observation of the $\mathcal{P}$-odd optical rotation in Xe, Hg, and I atoms was discussed. The techniques described in \cite{Boug14} is very close to what is necessary for the observation of the $\mathcal{P}$,$\mathcal{T}$-odd Faraday effect. In \cite{Chub18} an accurate evaluation of the $\mathcal{P}$,$\mathcal{T}$-odd Faraday effect oriented to the application of the techniques \cite{Boug14} was undertaken. Heavy atoms such as Cs, Tl, Pb, and Ra were chosen for these calculations. Heavy metal atoms such as Tl, Pb, Bi were considered as the most suitable objects for the observation of the $\mathcal{P}$-odd optical rotation in the old experiments \cite{Khrip91}. In the present paper we perform accurate calculations of the $\mathcal{P}$,$\mathcal{T}$-odd Faraday effect for Xe and Hg atoms considered in \cite{Boug14} as most suitable objects for the optical cavity experiments. Iodine atom that was also considered in \cite{Boug14} is not considered in the present paper since all suitable for the $\mathcal{P}$,$\mathcal{T}$-odd Faraday effect observation transitions lie in the short-wave ultraviolet region. Unlike \cite{Chub18} where the hyperfine structure of the atomic levels was ignored, in the present paper we evaluate the $\mathcal{P}$,$\mathcal{T}$-odd Faraday effect for the separate hyperfine sublevels. This corresponds to the experimental situation where the hyperfine structure is usually resolved.

\section{Theory}
The rotation angle $\psi$ for the polarization plane of the light propagating through the optically active medium with any type of birefringence (natural or $\mathcal{P}$-odd optical activity, ordinary or $\mathcal{P}$,$\mathcal{T}$-odd Faraday effect) is defined by the relation (see, for example \cite{Khrip91})
\begin{equation}
 \label{1}
\psi=\pi \frac{l}{\lambda} \text{Re} \left(n_+-n_-\right)
\end{equation}
where $l$ is the optical path length, $\lambda$ is the wavelength of the light and $n_{+(-)}$ are the refractive indices for the right (left) circularly polarized light. The refractive index for any resonant process in any atomic system is connected with the dynamic polarizability of this system $\alpha(\omega)$:
\begin{equation}
 \label{2}
n(\omega)\approx 1+2\pi\rho \alpha(\omega).
\end{equation}
Here $\rho$ is the atomic number density,
\begin{equation}
 \label{3}
\alpha_{\gamma JF} (\omega) = \frac{e^2}{3} \frac{1}{2F+1}\sum_{\gamma' J' F' M_F' M_F} \frac{\left|\langle \gamma JF M_F |  \bm{r}|\gamma' J'F'M_F'\rangle \right|^2}{E_{\gamma' J' F'}-E_{\gamma JF}-\omega -\frac{i}{2} \left(\Gamma_{\gamma JF}+\Gamma_{\gamma' J' F'}\right)}
\end{equation}
is the polarizability of atomic state $\gamma JF$, $J$ is the total electron angular momentum of an atom, $F$ denotes the total angular momentum of an atom including the nuclear spin (hyperfine structure level), $M_F$ denotes the projection of the total angular momentum. Summation in \Eq{3} is extended over the entire atomic spectra. In the resonance case only one term corresponding to the particular electron level $\gamma' J'$ and particular hyperfine sublevel $F'$ is retained. In the energy denominator $E_{\gamma JF}$ are the energies of the hyperfine sublevels of the electronic level $J$ and $\Gamma_{\gamma JF}$ are the corresponding widths. Polarization in \Eq{3} is averaged over the projection $M_F$ of the total momentum $F$ of an atom. 

In an external electric field the energy levels $E_{\gamma JF}$ begin to depend on $|M_F|$ and with the $\mathcal{P}$,$\mathcal{T}$-odd effects taken into account a sublevel with $|M_F|$ value is split in two levels with $M_F=\pm |M_F|$ having different energies, just like Zeeman structure. \Eq{3} then takes the form (in what follows we will consider only the resonant case, i.e. the transition $\gamma JF \rightarrow \gamma' J'F'$ between the hyperfine sublevels):
\begin{equation}
 \label{4}
\alpha_{\gamma JF\rightarrow \gamma' J'F'}^{\pm} (\omega) = \frac{e^2}{3} \frac{1}{2F+1}\sum_{M_F' M_F} \frac{\left|\langle \gamma JF M_F |  \bm{r}|\gamma' J'F'M_F'\rangle \right|^2}{\left(\omega^{(\pm)}_{\gamma JFM_F,\gamma' J' F'M_F'}-\omega\right) -\frac{i}{2} \left(\Gamma_{\gamma JFM_F}+\Gamma_{\gamma' J' F'M_F'}\right)},
\end{equation}
\begin{equation}
 \label{5}
\omega^{(+)}_{\gamma JFM_F, \gamma' J'F'M_F'}=E_{\gamma' J'F'M_F'}-E_{\gamma JFM_F},
\end{equation}
\begin{equation}
 \label{6}
\omega^{(-)}_{\gamma JFM_F, \gamma' J'F'M_F'}=E_{\gamma' J'F'\overline{M}_F'}-E_{\gamma JF \overline{M}_F}.
\end{equation}
Here $\overline{M}_F=-M_F$ and $E_{\gamma JFM_F}$ are the Stark split components of hyperfine sublevel $F$ of electronic level $\gamma J$. We are interested only in those components $M_F$, $M_F'$ which satisfy the condition
\begin{equation}
 \label{7}
M_F-M_F'=\pm 1.
\end{equation}
Only the transitions $\gamma JFM_F \rightarrow \gamma' J'F'M_F'$ which satisfy \Eq{7} correspond to the absorption of the right (left) circularly polarized photons and therefore exhibit the $\mathcal{P}$,$\mathcal{T}$-odd Faraday rotation.

The Stark component energies we present as
\begin{equation}
 \label{8}
E_{\gamma JF M_F}=E_{\gamma JF}^{(0)} + d_e \mathcal{E} \langle \gamma JF M_F| S^{\text{EDM}} |\gamma JF M_F \rangle 
\end{equation}
where $E_{\gamma JF}^{(0)}$ is the energy of the certain hyperfine sublevel in the absence of electric field, $\mathcal{E}$ is the magnitude of an external electric field strength and $\langle \gamma JF M_F| S^{\text{EDM}} |\gamma JF M_F \rangle$ is the shift of the linear Stark component caused by the existence of the $e$EDM. Expressions for the matrix elements $\langle \gamma JF M_F| S^{\text{EDM}} |\gamma JF M_F \rangle$ are given in Appendix A. In fact,
\begin{equation}
 \label{8a}
 \langle \gamma JF M_F| S^{\text{EDM}} |\gamma JF M_F \rangle \equiv R_d
\end{equation}
where $R_d$  is a dimensionless enhancement coefficient of the electron EDM in an atom.

In the case when the linear Stark shift is caused by the $\mathcal{P}$,$\mathcal{T}$-odd pseudoscalar-scalar electron-nucleus interaction $d_e$ in \Eq{8} should be replaced by $d_e^{\text{eqv}}$. A standard way of presenting such Stark component energies is as follows
\begin{equation}
 \label{8b}
E_{\gamma JF M_F}=E_{\gamma JF}^{(0)} + C_S \mathcal{E}  \langle \gamma JF M_F| S^{\text{SP}}|\gamma JF M_F \rangle
\end{equation}
where
\begin{equation}
 \label{8c}
 \langle \gamma JF M_F| S^{\text{SP}}|\gamma JF M_F \rangle  =  R_S.
\end{equation}
Here the constant $R_S$ interprets the EDM of an atom in terms of the dimensionless time-reversal-symmetry-violating electron-nucleon coupling parameter $C_S$. $R_S=d_e^{\text{eqv}}$ for $C_S=1$. For the $\mathcal{P}$,$\mathcal{T}$-odd pseudoscalar-scalar electron-nucleus interaction the corresponding effective operator $V^{\text{SP}}$ can be expressed in the following way \cite{San75},\cite{Gor79},\cite{Bon15}:
 \begin{equation}
 \label{8e}
 C_S \mathcal{E}  R_S= \mathcal{E}  \langle \gamma JF M_F| V^{\text{SP}}|\gamma JF M_F \rangle,
 \end{equation}
 \begin{equation}
 \label{8d}
 V^{\text{SP}}=Q_{\mathcal{P},\mathcal{T}}C_Si\frac{G_F}{\sqrt{2}}\gamma_0 \gamma_5 \rho(\bm{r}),
 \end{equation}
where $G_F$ is the Fermi-coupling constant and $Q_{\mathcal{P},\mathcal{T}}$ is the ``$\mathcal{P},\mathcal{T}$-odd charge of the nucleus'', in both models of the $\mathcal{P},\mathcal{T}$-odd electron-nucleus interaction \cite{Pos14},\cite{Chub16} $Q_{\mathcal{P},\mathcal{T}}=A$, where $A$ is the atomic number. $\rho(\bm{r})$ is the normalized nuclear density.

The second term in the right-hand side of \Eq{8} is extremely small compared to the first term, so we can expand \Eq{8}, Eqs \Br{5},\Br{6}, \Eq{4}, and \Eq{2} in powers of small parameter $d_e \mathcal{E}$, retaining only the first term of the expansion and inserting it into \Eq{1}. It is more convenient to do this after replacing the Lorentz profile in Eqs \Br{2}-\Br{4} by the Voigt profile, i.e. taking into account the Doppler broadening, the chaotic motion of atoms in a vapor (Maxwell distribution of velocities) and the collisional broadening. Under conditions most suitable for performing $\mathcal{P}$-odd or $\mathcal{P}$,$\mathcal{T}$-odd atomic experiments (atomic vapor density and temperature) the natural line width is smaller than the Doppler width but dominates over the collisional width. The real (dispersive) part of the refractive index $n(\omega)$ which in our case defines the $\mathcal{P}$,$\mathcal{T}$-odd Faraday rotation angle can be parametrized as (see, for example, \cite{Khrip91})
\begin{equation}
\label{9}
 \text{Re}\; n(u) \sim \text{Im} \; \mathcal{F} (u,v) \equiv g(u,v).
\end{equation}
The absorptive part is proportional to
\begin{equation}
\label{10}
\text{Im} \; n(u) \sim  \text{Re}\; \mathcal{F} (u,v) \equiv f(u,v).
\end{equation}
The function $\mathcal{F} (u,v)$ is defined as
\begin{equation}
\label{11}
\mathcal{F} (u,v) = \sqrt{\pi} e^{-(u+iv)^2} \left[ 1- \text{Erf} (-i(u+iv)) \right]
\end{equation}
where $\text{Erf}(z)$ is the error function, the variables $u$, $v$ are defined as
\begin{equation}
\label{12}
u=\frac{\Delta\omega}{\Gamma_D}
\end{equation}
and
\begin{equation}
\label{13}
v=\frac{\Gamma}{2\Gamma_D},
\end{equation}
respectively. Here $\Delta\omega$ is the frequency detuning, $\Gamma_D$ is the Doppler width and $\Gamma$ is the natural width. The Doppler width is equal to 
\begin{equation}
\label{14}
\Gamma_D=\omega_0 \sqrt{\frac{2k_B T}{Mc^2}}
\end{equation}
where $k_B$ is the Boltzmann constant, $T$ is the temperature in Kelvin, $M$ is the mass of an atom and $c$ is the speed of the light. 

The function $g(u,v)$ describes the behaviour of the optical rotation angle in case of optical activity (natural or $\mathcal{P}$-odd) in the vicinity of the resonance, this behaviour is depicted in Fig. 1(a). The function $f(u,v)$ describes the absorption line profile in the vicinity of the resonance, this line profile is presented in Fig. 1(b). In Fig. 1(c) the function $h(u,v)=\frac{dg}{du}$ is presented. All the pictures 1 (a,b,c) correspond to the case $v\ll 1$. The function $h(u,v)$ describes the rotation angle caused by the $\mathcal{P}$,$\mathcal{T}$-odd Faraday effect close to the resonance frequency. As it can be seen from Fig. 1(c) this function has two maxima (by absolute value): one maximum corresponding to the point of the resonance coinciding with maximum of absorption and another maximum off the resonance where absorption is small. This second maximum should allow to work off resonance when observing the ordinary Faraday effect or searching for the $\mathcal{P}$,$\mathcal{T}$-odd Faraday effect with the large optical path length.
\begin{figure}[h]
\begin{center}
\begin{minipage}[h]{0.3\linewidth}
\label{f:1a}
\center{\includegraphics[width=5 cm]{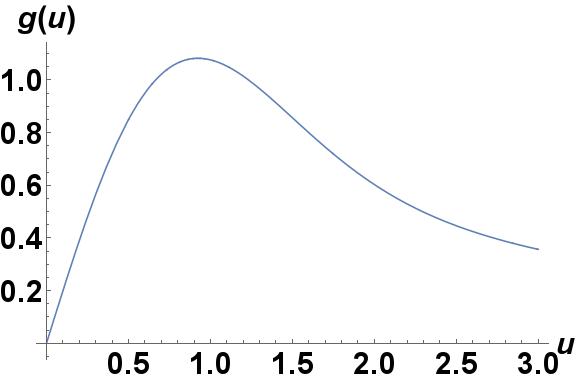}\\ (a)}
\end{minipage}
\hfill 
\begin{minipage}[h]{0.3\linewidth}
\center{\includegraphics[width=5 cm]{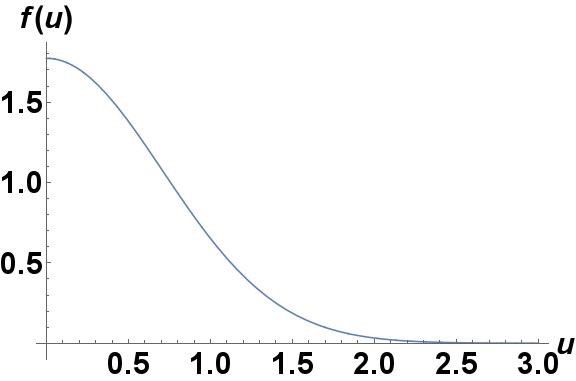} \\ (b)}
\label{f:1b}
\end{minipage}
\hfill
\begin{minipage}[h]{0.3\linewidth}
\center{\includegraphics[width=5 cm]{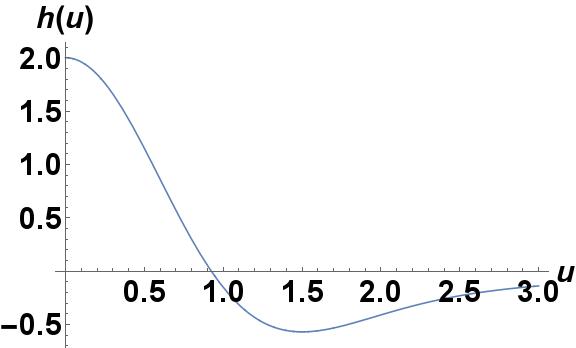} \\ (c)}
\label{f:1c}
\end{minipage}
\caption{\it Behaviour of the functions $g(u,v)$, $f(u,v)$ and $h(u,v)$ with $v\ll 1$ in the vicinity of the resonance. Fig. 1(a) represents the optical rotation angle (natural or $\mathcal{P}$-odd), Fig. 1(b) represents the inverse absorption length, Fig. 1(c) represents the rotation angle for the Faraday effect (ordinary or $\mathcal{P},\mathcal{T}$-odd).}
\end{center}
\end{figure}

Replacing now the Lorentz profile in \Eq{4} by the Voigt profile, using Eqs \Br{1} and \Br{2}, and expanding the result in terms of small parameter $d_e \mathcal{E}$ we find
\begin{eqnarray}
 \label{16}
\psi_{\gamma JF M_F,\gamma' J'F' M_F'} (\omega) & = & \frac{2\pi^2}{3} \frac{l}{\lambda} \rho e^2  \frac{1}{2F+1} \sum_{M_FM_F'}|\langle \gamma JF M_F |  \bm{r}| \gamma' J'F' M_F' \rangle|^2 \frac{h(u,v)}{\hbar\Gamma_D}  \nonumber
\\  & \times &   2d_e \mathcal{E} \frac{\langle \gamma' J'F' M_F'| S^{\text{EDM}} |\gamma' J'F' M_F' \rangle-\langle \gamma JF M_F| S^{\text{EDM}} |\gamma JF M_F \rangle}{\Gamma_D}.
\end{eqnarray}
The dependence of all the matrix elements in \Eq{16} on the quantum numbers $F$, $M_F$ is given in Appendix A.

Let us define the Faraday rotation signal $R(\omega)$ - the product of the Faraday rotation angle $\psi(\omega)$ and the light transmission function $T(\omega)$:
\begin{equation}
\label{17}
R_{\gamma JF M_F,\gamma' J'F' M_F'}(\omega)= \psi_{\gamma JF M_F,\gamma' J'F' M_F'}(\omega)T_{\gamma JF,\gamma' J'F' }(\omega).
\end{equation}
Transmission function does not depend on $M_F$, $M_F'$. Introducing again the Voigt profile the transmission function $T(\omega)$ can be presented as
\begin{equation}
\label{18}
T_{\gamma JF,\gamma' J'F' }=e^{-\rho l \sigma_{\gamma JF,\gamma' J'F' }f(u,v)}
\end{equation}
where $\sigma_{\gamma JF,\gamma' J'F' }$ is the absorption cross-section at the point of resonance
\begin{equation}
\label{19}
\sigma_{\gamma JF,\gamma' J'F' }=\frac{4\pi}{3\hbar c} \frac{\omega_0}{\Gamma_D}  \frac{e^2}{2F+1} \sum_{M_FM_F'}|\langle\gamma JF M_F |\bm{r} | \gamma' J'F' M_F'\rangle |^2. 
\end{equation}
The dependence of all the matrix elements in Eqs \Br{16}-\Br{19} on the hyperfine quantum numbers $F$, $M_F$ can be separated out (see Appendix A).

\section{Details of electronic structure calculations}

Direct use of \Eq{A7} corresponds to the so-called sum-over-states method. Formally, the summation in the equation should include all the excited states. In practice, only several contributions to this sum can be taken into account. However, it is possible to reformulate the problem: instead of explicit summation of the second order perturbation theory, one can calculate expression \Br{A7} as the mixed derivative of the energy with respect to the external electric field and $d_e$~\cite{Skripnikov:11a,Skripnikov:17a}. Note that in Ref.~\cite{Skripnikov:17a} the ``strategy I''  approach where one adds the interaction with the external electric field already at the self-consistent field stage of calculation was formulated. 

To calculate Stark shifts in the ground and excited electronic states of Xe and Hg we used the Relativistic Fock-Space coupled cluster with single and double cluster amplitudes method \cite{Visscher:01} to treat electron correlation effects.
In these calculations all electrons were included in the correlation calculation and the Dirac-Coulomb Hamiltonian was employed. The uncontracted Dyall's AETZ \cite{Dyall:07,Dyall:12,Dyall:09} basis sets were used in the calculations augmented by several diffuse functions of $s$-, $p$- and $d$- types.

E1  transition matrix elements were calculated using the multireference linear response coupled cluster with single and double amplitudes (CCSD) method \cite{Kallay:3,Kallay:5,Kallay:2} for Xe and single reference linear response CCSD for Hg. For these calculations the Dyall's AEDZ basis sets \cite{Dyall:07,Dyall:12,Dyall:09} with additional diffuse functions were used. $1s..3d$ electrons were excluded from the correlation treatment of Xe and $1s..4f5s$ electrons were excluded from the correlation treatment of Hg in the case of E1  transition matrix elements .

Electronic calculations were performed within the {\sc dirac12} \cite{DIRAC12} and {\sc mrcc} \cite{MRCC2013} codes. Matrix elements of operators of E1 transitions and $\mathcal{P}$,$\mathcal{T}$-odd interactions were calculated using code developed in Refs.~\cite{Skripnikov:17a,Skripnikov:16b,Petrov:17b}. 

Uncertainty of the enhancement factors can be estimated by 15\%.

\begin{table} [h!]
\caption{Enhancement dimensionless coefficients $R_d$ for the electron EDM effect and factors $R_S$ ($R_S=d_e^{\text{eqv}}(C_S=1)$) for the $\mathcal{P}$,$\mathcal{T}$-odd electron-nucleus interaction effect for Xe and Hg certain hyperfine sublevels of the electronic states under consideration} \centering
\begin{tabular}{ccccc}
\hline\hline  Atom & Configuration & Term  & $R_d$ & $R_S\times10^{18} $,   \\
                   &               &       &  &  $e$ cm \\
\hline Xe & $(^2P_{3/2}^0) 6s$  & $^2[3/2]^0_2$, $F$=$\frac{3}{2}$, $M_F$=$-\frac{1}{2}$ & $-34$ & $-0.209$  \\
\,  &  $(^2P_{3/2}^0)6p$  & $^2[1/2]_1$, $F$=$\frac{1}{2}$, $M_F$=$\frac{1}{2}$  & $-32$ & $-0.221$  \\
\hline  Hg & $6s6p$ & $^3P_1$, $F$=$\frac{1}{2}$ ,$M_F$=$-\frac{1}{2}$  & 285 &  3.31  \\
  \,    & $6s7s$  & $^3S_1$, $F$=$\frac{1}{2}$, $M_F$=$\frac{1}{2}$  & 595 & 6.87 \\
 \,   & $6s^2$ & $^1S_0$, $F$=$\frac{1}{2}$, $M_F$=$\frac{1}{2}$  & 0 & 0 \\
\hline \hline
\end{tabular}
\label{table:1}
\end{table}

\section{Results of calculations for Xe and Hg atoms}
\subsection{Xe atom}

\begin{figure}[h!]
\begin{center}
\includegraphics[width=8.0 cm]{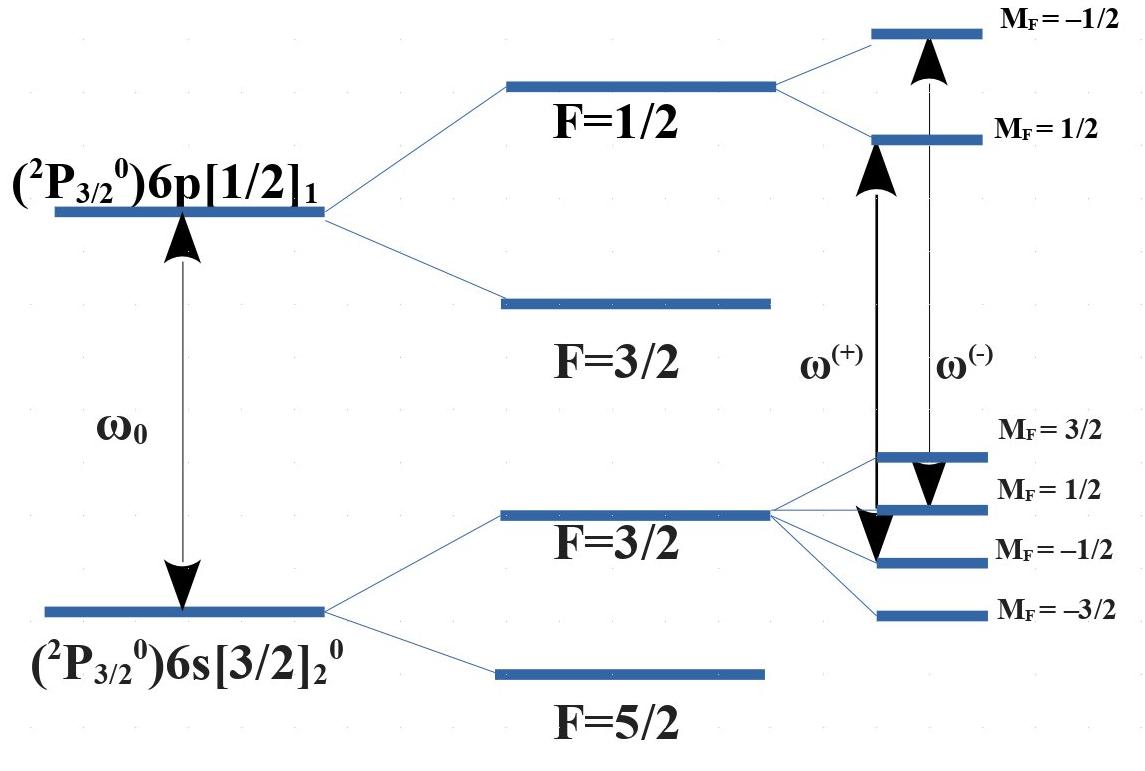}
\end{center}
\caption{\label{f:2} { The scheme of the hyperfine and the linear Stark splitting for the E1 transition in Xe atom}}
\end{figure}

\begin{figure}[h!]
\begin{center}
\includegraphics[width=10.0 cm]{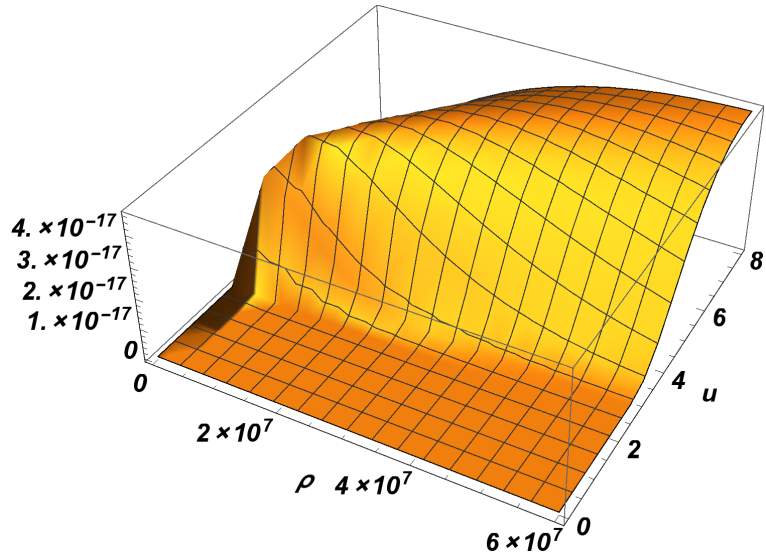}
\end{center}
\caption{\label{f:3} {Dependence of the $\mathcal{P}$,$\mathcal{T}$-odd Faraday signal $R$ (in rad) on dimensionless detuning $u$ and on vapor density $\rho$ (in cm$^{-3}$) for the E1 transition $(^2P_{3/2}^0) 6s [3/2]^0_2 \rightarrow (^2P_{3/2}^0) 6p [1/2]_1$  in Xe atom. The optical path length $l$ is assumed to be equal to 100 km}}
\end{figure}

In the case of Xe atom we consider the E1 transition from the metastable $(^2P_{3/2}^0) 6s [3/2]^0_2$ state to the excited $(^2P_{3/2}^0) 6p [1/2]_1$ state. We choose the isotope $^{129}$Xe with the nuclear spin $I=1/2$. Transition wavelength is $\lambda=980$ nm (the experimental transition energy is $\Delta E=1/\lambda=10202$ cm$^{-1}$ \cite{Hum70}). Our calculated transition energy $\Delta E=10224$ cm$^{-1}$ is in a very good agreement with the experiment. The calculation of the squared value of the reduced E1 matrix element yields $| \langle(^2P_{3/2}^0) 6s [3/2]^0_2 || r ||(^2P_{3/2}^0) 6p [1/2]_1 \rangle|^2= 42.85$ a.u.$^2$. A scheme of the hyperfine and the linear Stark splitting of the levels for this transition is given in Fig. \ref{f:2}.
The population of the lower metastable level can be obtained with the laser pumping \cite{Boug14}.
The evaluation of $(\omega^{(+)}-\omega^{(-)})$ for the $e$EDM effect according to Eqs \Br{5},\Br{6},\Br{8} results in
\begin{eqnarray}
\label{20}
(\omega^{(+)}-\omega^{(-)})&=&2\left(R_d\left( (^2P_{3/2}^0)6p [1/2]_1, F\!=\! 1/2,M_F\!=\! 1/2 \right)-R_d \left((^2P_{3/2}^0) 6s [3/2]^0_2,F\!=\!3/2,M_F\!=\! -1/2\right)\right)d_e \mathcal{E}\nonumber
\\
   &=& 2\left(-32-(-34)\right)d_e \mathcal{E}=4 \times d_e \mathcal{E}.
\end{eqnarray}
Note, that with the claimed uncertainty the difference \Eq{20} can be close to zero, however, present estimations are correct by an order of magnitude for other hyperfine transitions.
See Table I for the enhancement coefficients $R_d$ for the electron EDM effect and the factors $R_S$ for the $\mathcal{P}$,$\mathcal{T}$-odd electron-nucleus interaction effect. In what follows the estimates of the  $\mathcal{P}$,$\mathcal{T}$-odd signal $R$ are made for the electron EDM effect assuming $C_S=0$ and $d_e= 1.1 \times 10^{-29}$ $e$ cm (the bound established in the experiment with the ThO molecule \cite{ACME18}). The similar estimates also can be made for the $\mathcal{P}$,$\mathcal{T}$-odd electron-nucleus interaction effect assuming $d_e=0$ and $C_S=7.3 \times 10^{-10}$ (the bound established in the experiment with the ThO molecule \cite{ACME18}).
For an external electric field we set $\mathcal{E}=10^5$ V/cm \cite{Reg02}. Assuming the room temperature $T\sim300$ K and employing the transition frequency value $\omega_0=2\times 10^{15}$ s$^{-1}$, according to \Eq{14} we obtain the characteristic value for the Doppler width $\Gamma_D =6.5\times 10^{-7} \omega_0\approx 1.3 \times 10^9$ s$^{-1}$. The natural line width for the chosen transition is $\Gamma=2.6 \times 10^7$ s$^{-1}$. Using the optical path $l=100$ km \cite{Boug14} our calculation according to Eqs \Br{16}-\Br{19} gives the dependence $R(u,\rho)$ depicted in Fig.\ref{f:3}.
Then it follows from Fig.\ref{f:3} that the optimal number density of Xe atom vapors for the above conditions is $\rho_{\text{opt}}=3\times 10^{7}$ cm$^{-3}$ and $u_{\text{opt}}\approx 5$ which gives the maximum value of the effect.
Then
\begin{equation}
\label{21}
R_{\text{max}} (l=100 \, \text{km}) \approx 4.0 \times 10^{-17} \, \text{rad}
\end{equation}
for the observation of the electron EDM of the order $d_e\sim  10^{-29}$ $e$ cm. The $\mathcal{P}$,$\mathcal{T}$-odd Faraday effect is proportional to the difference between the enhancement coefficients for the states (which satisfy the condition \Eq{7}) between which the transition is considered  ($R\sim |\Delta R_d|$). Such a small value of the effect is mainly caused by this relation (in case of Xe atom  $R\sim |\Delta R_d|=2$). A real observable quantity in the experiment is the rotation angle of the light polarization plane but its value is limited by absorption. Analysing \Eq{21} and using a more familiar for experimentalists quantity $\rho l= 3\times 10^{14}$ cm$^{-2}$ (which referred to as the column density) for Xe one can evaluate the maximum rotation angle $\psi_{\text{max}} \sim 2 \times 10^{-16}$ rad. This result shows that the best possible estimate for the $e$EDM with  ICAS maximum modern sensitivity achievement ($\sim 10^{-13}$ rad \cite{Dur10}) would be still 3 orders of magnitude above the value quoted in \cite{ACME18}.  

\subsection{Hg atom}
\begin{figure}[h!]
\begin{center}
\includegraphics[width=8.0 cm]{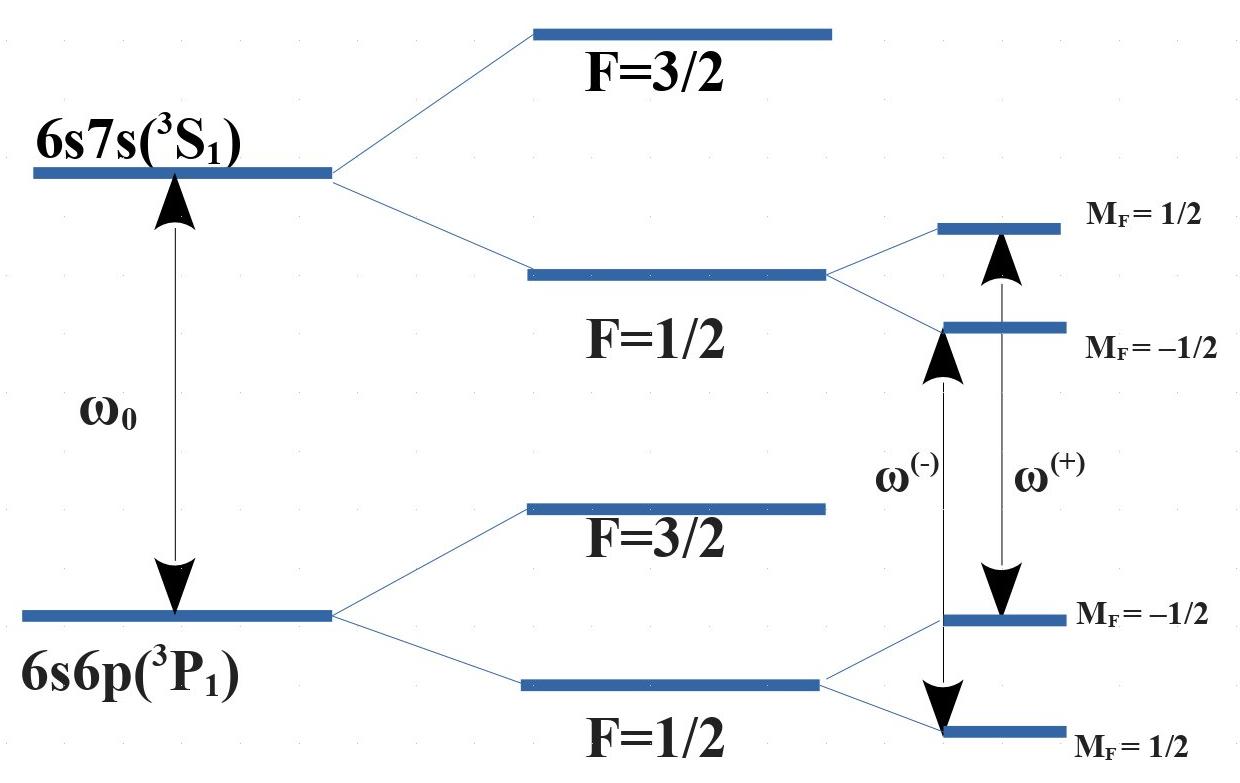}
\end{center}
\caption{\label{f:4} {The scheme of the hyperfine and the linear Stark splitting for the $6s6p (^3P_1) \rightarrow 6s7s(^3S_1)$ E1 transition in $^{199}$Hg atom}}
\end{figure}

\begin{figure}[h!]
\begin{center}
\includegraphics[width=10.0 cm]{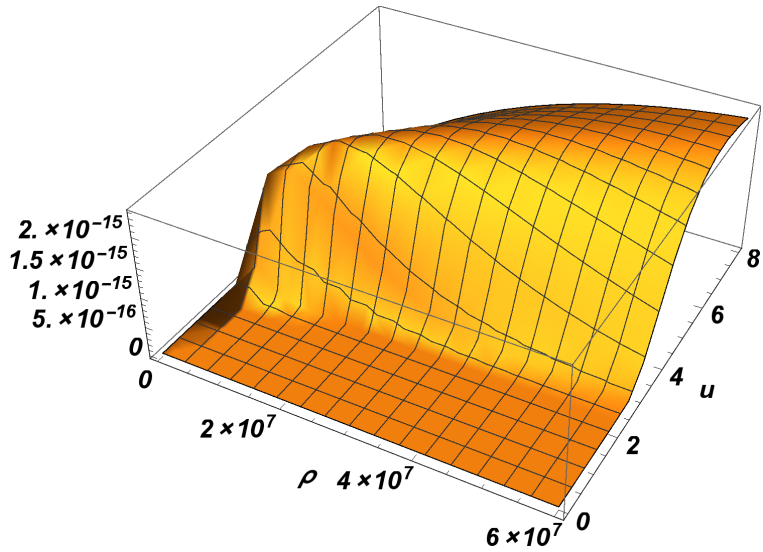}
\end{center}
\caption{\label{f:5} {Dependence of the $\mathcal{P}$,$\mathcal{T}$-odd Faraday signal $R$ (in rad) on dimensionless detuning $u$ and on vapor density $\rho$ (in cm$^{-3}$) for the  $6s6p (^3P_1) \rightarrow 6s7s(^3S_1)$ E1 transition in Hg atom. The optical path length $l$ is assumed to be equal to 100 km}}
\end{figure}

\begin{figure}[h!]
\begin{center}
\begin{minipage}[h!]{0.49\linewidth}
\center{\includegraphics[width=7 cm]{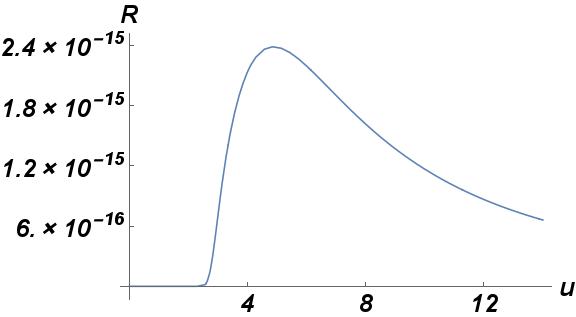}\\ (a)}
\end{minipage}
\hfill 
\begin{minipage}[h!]{0.49\linewidth}
\center{\includegraphics[width=7 cm]{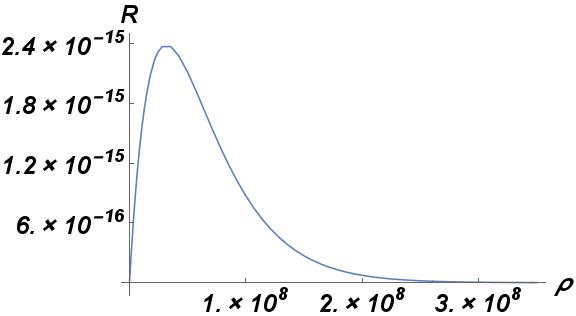} \\ (b)}
\end{minipage}
\caption{\it (a) Behaviour of the $R(u,\rho_{\text{opt}})$ projection of Fig.\ref{f:5} (in rad) assuming fixed number density $\rho_{\text{opt}}=4\times 10^{7}$ cm$^{-3}$, and (b) behaviour of the $R(u_{\text{opt}},\rho)$ projection of the Fig.\ref{f:5} (in rad) assuming fixed dimensionless detuning $u_{\text{opt}}=5$ ($\rho$ in cm$^{-3}$)}
\label{f:6}
\end{center}
\end{figure}

In the case of Hg atom we consider two E1 transitions for the isotope $^{199}$Hg with the nuclear spin $I=1/2$. The first one is from the metastable $6s6p (^3P_1)$ state to the excited $6s7s(^3S_1)$ state. The wavelength for this transition is $\lambda=436$ nm (the experimental transition energy is $\Delta E=1/\lambda=22938$ cm$^{-1}$ \cite{Sal06}). Our calculated transition energy $\Delta E=21028$ cm$^{-1}$ is in a very good agreement with the experiment. The calculation of the squared value of the reduced E1 matrix element yields $|\langle 6s6p (^3P_1) || r || 6s7s(^3S_1) \rangle | ^2=9.83$ a.u.$^2$. A scheme of the hyperfine and linear Stark splitting is given in Fig. \ref{f:4}. 
The population of the lower metastable level can be obtained with the laser pumping \cite{Boug14}.
The evaluation of $(\omega^{(+)}-\omega^{(-)})$ for the $e$EDM effect for this case according to the formulas Eqs \Br{5},\Br{6},\Br{8} results in
\begin{eqnarray}
\label{22}
(\omega^{(+)}-\omega^{(-)})&=&2\left(R_d\left(6s7s(^3S_1), F\!=\! 1/2,M_F\!=\! 1/2 \right)-R_d \left(6s6p (^3P_1),F\!=\! 1/2,M_F \!=\! -1/2\right)\right)d_e \mathcal{E}\nonumber
\\
   &=& 2\left(595-285 \right)d_e \mathcal{E}=620 \times d_e \mathcal{E}
\end{eqnarray}
(Also see Table I for the enhancement coefficients $R_d$ for the electron EDM effect and the factors $R_S$ for the $\mathcal{P}$,$\mathcal{T}$-odd electron-nucleus interaction effect). For an external electric field we again set $\mathcal{E}=10^5$ V/cm \cite{Reg02}. The natural line width for the chosen transition is $\Gamma=1.0\times 10^8$ s$^{-1}$. Assuming the room temperature $T\sim300$ K and employing the transition frequency value $\omega_0=4\times 10^{15}$ s$^{-1}$, according to \Eq{14} we obtain the characteristic value for the Doppler width $\Gamma_D =5.2\times 10^{-7} \omega_0\approx 2 \times 10^9$ s$^{-1}$. Using the optical path $l=100$ km \cite{Boug14} our calculation according to Eqs \Br{16}-\Br{19} gives the dependence $R(u,\rho)$ depicted in Fig.\ref{f:5}.
Then it follows from Fig.\ref{f:5} that the optimal number density of Hg atom vapors for the above conditions is $\rho_{\text{opt}}=4\times 10^{7}$ cm$^{-3}$ and $u_{\text{opt}}\approx 5$ which gives the maximum value of the effect. $R(u,\rho_{\text{opt}})$ and $R(u_{\text{opt}},\rho)$ projections of Fig.\ref{f:5} are presented in Fig.\ref{f:6} (a) and (b), respectively.
Then
\begin{equation}
\label{23}
R_{\text{max}} (l=100 \, \text{km}) \approx 2.4 \times 10^{-15} \, \text{rad}
\end{equation}
for the observation of the electron EDM of the order $d_e\sim  10^{-29}$ $e$ cm.
Analysing \Eq{23} and using the column density value of  $\rho l= 4\times 10^{14}$ cm$^{-2}$ for this transition in Hg one can evaluate the maximum rotation angle $\psi_{\text{max}} \sim  10^{-14}$ rad.

\begin{figure}[h!]
\begin{center}
\includegraphics[width=8.0 cm]{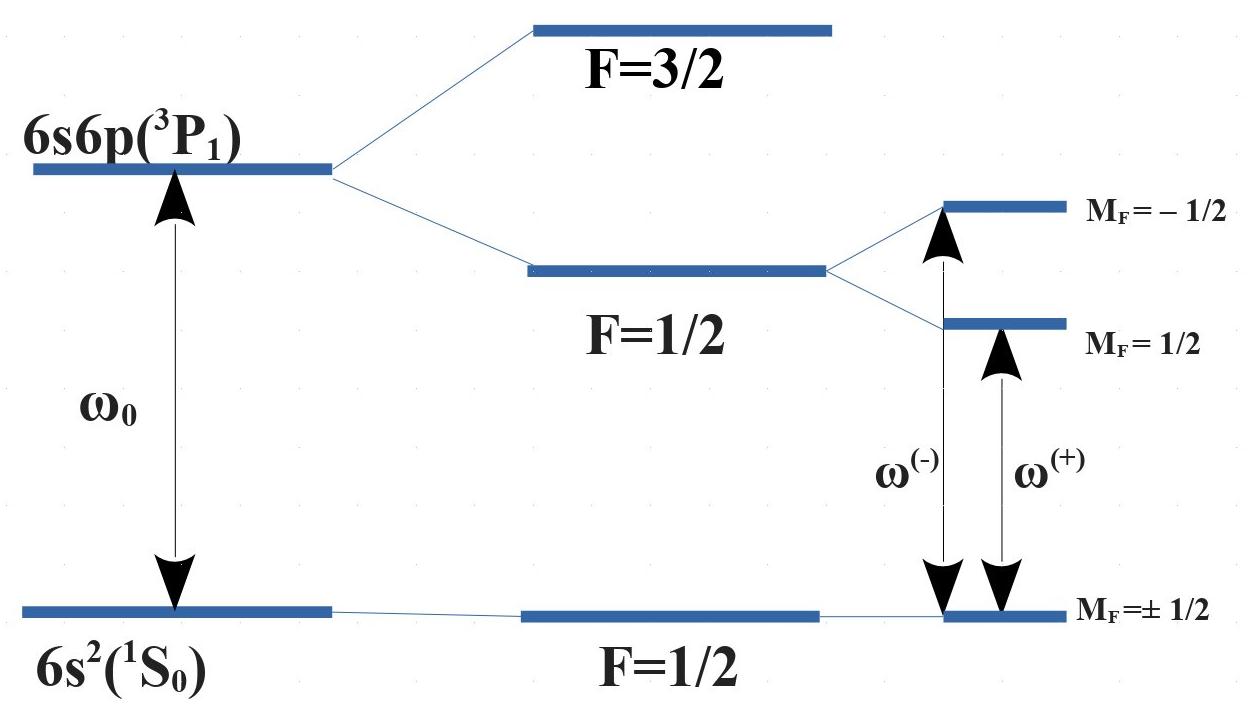}
\end{center}
\caption{\label{f:7} {The scheme of the hyperfine and the linear Stark splitting for the $6s^2 (^1S_0)\rightarrow 6s6p (^3P_1)$ E1 transition in $^{199}$Hg atom}}
\end{figure}

\begin{figure}[h!]
\begin{center}
\includegraphics[width=10.0 cm]{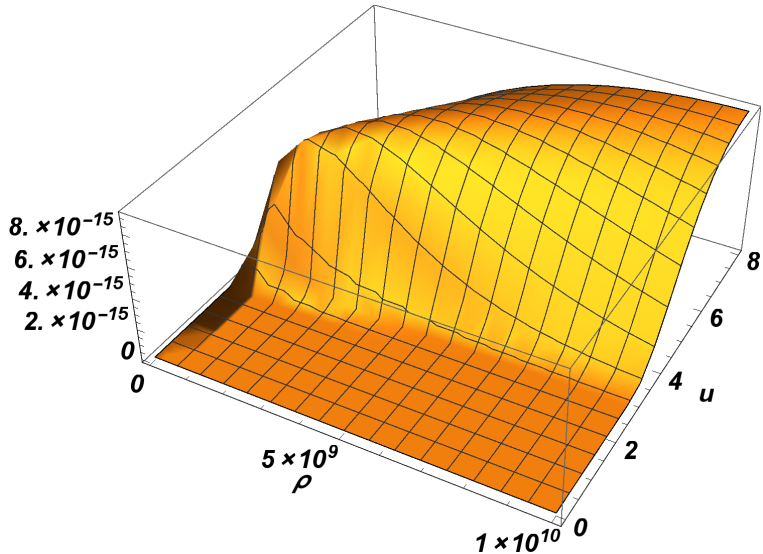}
\end{center}
\caption{\label{f:8} {Dependence of the $\mathcal{P}$,$\mathcal{T}$-odd Faraday signal $R$ (in rad) on dimensionless detuning $u$ and on vapor density $\rho$ (in cm$^{-3}$) for the  $6s^2 (^1S_0)\rightarrow 6s6p (^3P_1)$ E1 transition in Hg atom. The optical path length $l$ is assumed to be equal to 100 km}}
\end{figure}

The second E1 transition is from the ground $6s^2 (^1S_0)$ to the metastable $6s6p (^3P_1)$ state with the wavelength  $\lambda=254$ nm (the experimental transition energy is $\Delta E=1/\lambda=39412$ cm$^{-1}$ \cite{Sal06}). Our calculated transition energy $\Delta E=39806$ cm$^{-1}$ is in a very good agreement with the experiment. The calculation of the squared value of the reduced E1 matrix element yields $|\langle 6s6p (^3P_1) || r ||6s^2 (^1S_0) \rangle | ^2=0.42$ a.u.$^2$. A scheme of the hyperfine and linear Stark splitting is given in Fig. \ref{f:7}. 
The evaluation of $(\omega^{(+)}-\omega^{(-)})$ for the $e$EDM effect for this case according to the formulas Eqs \Br{5},\Br{6},\Br{8} results in
\begin{eqnarray}
\label{24}
(\omega^{(+)}-\omega^{(-)})&=&2\left(R_d\left(6s6p (^3P_1),F\!=\! 1/2,M_F \!=\! 1/2\right)-R_d \left(6s^2 (^1S_0),F\!=\! 1/2,M_F \!=\! -1/2\right)\right)d_e \mathcal{E}\nonumber
\\
   &=& 2\left(-285 \right)d_e \mathcal{E}=-570 \times d_e \mathcal{E}.
\end{eqnarray}
Employing the transition frequency value $\omega_0=7.4\times 10^{15}$ s$^{-1}$, according to \Eq{14} we obtain the characteristic value for the Doppler width $\Gamma_D =5.2\times 10^{-7} \omega_0\approx 3.7 \times 10^9$ s$^{-1}$.  The natural line width for the chosen transition is $\Gamma=2.0 \times 10^7$ s$^{-1}$. Using the optical path $l=100$ km \cite{Boug14} our calculation according to Eqs \Br{16}-\Br{19} gives the dependence $R(u,\rho)$ depicted in Fig.\ref{f:8}.
Then it follows from Fig.\ref{f:8} that the optimal number density of Hg atom vapors for the above conditions is $\rho_{\text{opt}}=4\times 10^{9}$ cm$^{-3}$ and $u_{\text{opt}}\approx 5$ which gives the maximum value of the effect. Then
\begin{equation}
\label{25}
R_{\text{max}} (l=100 \, \text{km}) \approx 1.0 \times 10^{-14} \, \text{rad}
\end{equation}
for the observation of the electron EDM of the order $d_e\sim  10^{-29}$ $e$ cm. The uncertainty of predicted value for the $\mathcal{P}$,$\mathcal{T}$-odd Faraday signal is defined by the uncertainty of the electronic structure calculations made in Section III and is about 15\%. Analysing \Eq{25} for this transition in Hg one can evaluate the maximum rotation angle $\psi_{\text{max}} \sim  10^{-13}$ rad. 

The results for Hg show that with the best sensitivity achievements of the modern ICAS ($\sim 10^{-13}$ rad \cite{Dur10}) the $\mathcal{P}$,$\mathcal{T}$-odd Faraday experiment with Hg atom could give the same upper bound for $e$EDM as already quoted value in \cite{ACME18}. Note that the results for the $\mathcal{P}$,$\mathcal{T}$-odd Faraday rotation signal given in \cite{Chub18} were overestimated for E1 transitions and should be at the same level as the results given here for Hg atom.

\section{Conclusions}
The main result of our studies reported in this paper is that the Hg atom is as favorable for the observation of the $\mathcal{P}$,$\mathcal{T}$-odd Faraday effect in ICAS experiments as the Cs,Tl,Pb, and Ra atoms studied earlier in \cite{Chub18} but unlike the latter ones is more suitable for the ICAS experiments of the type discussed in \cite{Boug14},\cite{Boug12} for observation of the $\mathcal{P}$-odd optical activity. To give the same upper bound for $e$EDM as is already reached in experiments with the electron spin precession in electric field \cite{ACME18} it would be necessary to use the maximum modern results in ICAS sensitivity $10^{-13}$ rad \cite{Dur10}. Recently several suggestions were made how to improve further the accuracy achieved in the observation of $e$EDM. The main problem in ACME experiments is the relatively short coherence time (few ms), i.e. the time of interaction for the molecule in the molecular beam with an external electric field. For the molecular ions which can be trapped in magnetic storage rings the coherence time becomes as large as $\sim 1$ s. The best experiment with molecular ion HfF$^{+}$ was reported in \cite{Cair17}. However the charged particles (molecular ions) are less robust with respect to systematic errors. Also there was a suggestion to trap neutral molecules using the laser cooling \cite{Koz17}. The polyatomic molecules (for example, YbOH) were considered as the most suitable candidates for this cooling. What concerns the $\mathcal{P}$,$\mathcal{T}$-odd Faraday experiment, the coherence time is limited only by the optical path length. The main disadvantage of the proposed $\mathcal{P}$,$\mathcal{T}$-odd Faraday experiments with atoms is the necessity to use very high electric field ($10^5$ V/cm) for obtaining acceptable rotation angle. Such a field can be easily produced within a small volume of the size 1 cm. To obtain such a field within all the cavity of 1 m long is a serious technical problem which we do not discuss in the present paper. One possible way to avoid this difficulty is to perform the ICAS $\mathcal{P}$,$\mathcal{T}$-odd Faraday experiment with diatomic molecules. For heavy diatomic molecules the electric field necessary to reach the same $\mathcal{P}$,$\mathcal{T}$-odd Faraday effect as in heavy atoms may be much weaker \cite{ACME13,ACME18,Chub14}. The work on the investigation of ICAS $\mathcal{P}$,$\mathcal{T}$-odd Faraday effect in heavy diatomic molecules is underway and is planned to be our next communication.

\section*{Acknowledgements}

In preparing the paper and the calculations of the $\mathcal{P}$,$\mathcal{T}$-odd Faraday signals, optimal parameters for the experiments and E1 amplitudes were supported by the Russian Science Foundation grant 17-12-01035. Calculation of the enhancement factors were supported by the Foundation for the advancement of theoretical physics and mathematics ``BASIS'' grant according to the research projects No.~18-1-3-55-1 and No.~17-15-577-1 and also by the President of Russian Federation Grant No. MK-2230.2018.2. The authors are grateful to Dr. T. Peter Rakitzis and Dr. Lykourgos Bougas for useful comments.
\section*{Appendix A: Separation of the dependence on the hyperfine quantum numbers $F$, $M_F$ in the matrix elements}

We start with the matrix element $\langle \gamma JF M_F |  \bm{r}| \gamma' J'F' M_F' \rangle$ in \Eq{16}. These matrix elements enter in \Eq{16} in the form of a scalar product of two irreducible tensor operators
\begin{equation*}
\label{A1}
\frac{1}{2F+1} \sum_{M_{F} M_{F}'} \sum_{q=0,\pm 1} \left(-1\right)^q \langle \gamma JF M_{F} |  r^1_q| \gamma' J'F' M_{F}' \rangle\langle \gamma' J'F' M_{F}' |  r^1_{\overline{q}}| \gamma JF M_{F} \rangle\tag{\text{A1}}
\end{equation*}
where $P_{\alpha}^a$ denotes the component $\alpha$ of the irreducible tensor of the rank $a$. Thus, we have to evaluate the matrix element of irreducible tensor operator depending on the variables of one subsystem (electron) with the wave functions depending on the variables of two subsystems (electron$+$nucleus). The application of the Wigner-Eckart theorem in this case results \cite{Var88}
 \begin{align*}
 \label{A2}
\langle \gamma_1' j_1' \gamma_2' j_2' j'm'|P_{\alpha}^a|  \gamma_1 j_1 \gamma_2 j_2 jm\rangle 
&=
\delta_{\gamma_2\gamma_2'} \delta_{j_2j_2'} \left(-1\right)^{j+j_1'+j_2-a}  \Pi_j C_{jma\alpha}^{j'm'}
\\  
&\times  
{\begin{Bmatrix}
   j_1 & j_2 & j \\
   j' & a & j_1' 
  \end{Bmatrix}}  \langle \gamma_1' j_1'||P^a|| \gamma_1 j_1 \rangle. \tag{\text{A2}}
\end{align*}
Here $j_1$, $j_2$ are the angular momenta for two subsystems, $j$, $m$ are the total angular momentum and its projection. The notations from \cite{Var88} for the Clebsh-Gordan coefficients, $6j$-symbols and reduced matrix elements are employed.
\begin{equation*}
\label{A3}
\Pi_{ab\dots c}=\sqrt{(2a+1)(2b+1)\dots (2c+1)} \tag{\text{A3}}
\end{equation*}
In our case $jm=FM_F$, $j'm'=F'M_F'$, $j_1=J$, $j_1'=J'$, $j_2=j_2'=I$, $a=1$, $\alpha=q$ for one matrix element and $jm=F'M_F'$, $j'm'=FM_F$, $j_1=J'$, $j_1'=J$, $j_2=j_2'=I$, $a=1$, $\alpha=\overline{q}=-q$ for another matrix element where $I$ is the nuclear spin. Then \Eq{A1} looks like
 \begin{align*}
 \label{A4}
\frac{1}{2F+1} 
& \sum_{M_{F} M_{F}'q} 
\left(-1\right)^{q+F+F'+J+J'+2I-2} \sqrt{(2F+1)(2F'+1)}  \times
C_{FM_{F}1\overline{q}}^{F'M_{F}'}C_{F'M_{F}'1q}^{FM_{F}}
\\  
&\times  
{\begin{Bmatrix}
   J & I & F \\
   F' & 1 & J' 
  \end{Bmatrix}} 
  {\begin{Bmatrix}
   J' & I & F' \\
   F & 1 & J 
  \end{Bmatrix}}
   |\langle \gamma J||r^1|| \gamma' J' \rangle|^2. \tag{\text{A4}}
\end{align*}
Summation over $M_{F}, M_{F}',q$ in \Eq{A4} can be reduced to the factor
\begin{equation*}
\label{A5}
 \frac{1}{2F+1}\sum_{M_{F} M_{F}'q}  \left(-1\right)^{q} C_{FM_{F}1\overline{q}}^{F'M_{F}'}C_{F'M_{F}'1q}^{FM_{F}}=\left(-1\right)^{F+F'+s} \sqrt{\frac{2F'+1}{2F+1}},   \tag{\text{A5}}
\end{equation*}
where $s= 2F'\;  (\text{mod} \; 2)$. 
Then the expression \Eq{A1} takes the form 
 \begin{align*}
 \label{A6}
\left(-1\right)^{J+J'+2I+s} (2F'+1){\begin{Bmatrix}
   J & I & F \\
   F' & 1 & J' 
  \end{Bmatrix}} 
  {\begin{Bmatrix}
   J' & I & F' \\
   F & 1 & J 
  \end{Bmatrix}}
   |\langle \gamma J||r^1|| \gamma' J' \rangle|^2. \tag{\text{A6}}
\end{align*}

Next we consider the matrix element $\langle \gamma JF M_F| S^{\text{EDM}} |\gamma JF M_F \rangle$ in \Eq{8}. This matrix element looks like \cite{Chub17}, \cite{Chub18}
\begin{align*}
 \label{A7}
d_e \mathcal{E}\langle  \gamma JF M_F |S^{\text{EDM}}| \gamma JF M_F\rangle 
&= -d_e \langle\gamma JF M_F |(\gamma_0-1) \bm{\mathcal{E}}\bm{\Sigma} |\gamma JF M_F \rangle
 \\ 
 &+  d_e e \bm{\mathcal{E}}  \sum_{ \gamma' J'F' M_F'} \Biggl\{ \frac{\langle  \gamma JF M_F |\bm{r}| \gamma' J'F' M_F' \rangle \langle  \gamma' J'F' M_F' |(\gamma_0-1) \bm{\mathcal{E}_c}\bm{\Sigma} |  \gamma JF M_F \rangle}{{E}_{ \gamma' J'F'}-{E}_{ \gamma JF}}
\\
&+ \frac{\langle  \gamma JF M_F |(\gamma_0-1) \bm{\mathcal{E}_c}\bm{\Sigma} |  \gamma' J'F' M_F'\rangle  \langle   \gamma' J'F' M_F'|\bm{r}| \gamma JF M_F \rangle }{{E}_{ \gamma' J'F'}-{E}_{\gamma JF}} \Biggr\}.   \tag{\text{A7}}
\end{align*}
Here $\bm{\mathcal{E}_c}$ is the strength of the Coulomb field of the nucleus and other electrons, $e$ is the electron charge (by modulus), $\bm{r}$ is the electron radius-vector, $r=|\bm{r}|$; $\gamma_0$,$\bm{\Sigma}$ are the Dirac matrices. \Eq{A7} is written for an atom with one valence electron. In case of several valence electrons the one-electron operators in the matrix elements in \Eq{A7} should be replaced by the sums of one-electron operators for all the electrons. In the latter case the wave functions in the matrix elements in \Eq{A7} should be the many-electron ones and the quantum numbers $\gamma J$ should belong to the whole atom. In this way the electron correlation within any approximation can be taken into account.
Note also, that according to \cite{John86} within the Dirac-Coulomb Hamiltonian one can use
alternative expression for the $e$EDM interaction in \Eq{A7}:
\begin{align*}
  V^{\textit{e}{\rm EDM}}= d_e\frac{2i}{e\hbar}c\gamma^0\gamma^5\bm{p}^2, \tag{\text{A8}}
 \label{A8}
\end{align*}
where $\bm{p}$ is the electron momentum operator. The advantage of such form of the interaction is that it is written in the one-electron form.

The first term in the right-hand side of \Eq{A7} usually gives a negligible contribution, so that we will consider only the second and the third terms. To demonstrate the separation of the hyperfine quantum numbers $FM_F$ in the matrix element \Eq{A7} we consider an atom with one valence electron in the one-electron approximation.

We present the matrix element $d_e \mathcal{E} \langle \gamma JF M_F| S^{\text{EDM}} |\gamma JF M_F \rangle$ in the form
\begin{equation*}
\label{A9}
 d_e \mathcal{E} \langle \gamma JF M_F| S^{\text{EDM}} |\gamma JF M_F \rangle = d_e \langle \gamma JF M_F|\bm{\mathcal{E}} \bm{\Sigma}^{\text{EDM}} |\gamma JF M_F \rangle \tag{\text{A9}}
\end{equation*}
where $\bm{\mathcal{E}}$ is an external electric field and vector $\bm{\Sigma}^{\text{EDM}}$ is defined by \Eq{A7}. Let the field $\bm{\mathcal{E}}$ be oriented along $z$ axis then $\mathcal{E}_x=\mathcal{E}_y=0, \mathcal{E}_z=\mathcal{E}$ and
\begin{equation*}
\label{A10}
 d_e \mathcal{E} \langle \gamma JF M_F| S^{\text{EDM}} |\gamma JF M_F \rangle = d_e \mathcal{E}\langle \gamma JF M_F| \Sigma_0^{\text{EDM},1} |\gamma JF M_F \rangle \tag{\text{A10}}
\end{equation*}
where $\Sigma_0^{\text{EDM},1}$ is zero-component of the irreducible tensor $\Sigma_q^{\text{EDM},1}$ of the rank 1 corresponding to the vector $\bm{\Sigma}^{\text{EDM}}$. Now we can apply again the general formula \Eq{A2} since the tensor $\Sigma_0^{\text{EDM},1}$ depends only on the variables of the electron subsystem in the total atomic system (electrons+nucleus):
 \begin{align*}
 \label{A11}
d_e \mathcal{E}\langle \gamma JF M_F| \Sigma_0^{\text{EDM},1} |\gamma JF M_F \rangle 
&= \left(-1\right)^{F+J+I-1} \sqrt{2F+1} C_{FM_F10}^{FM_F}
\\  
&\times  
{\begin{Bmatrix}
   J & I & F \\
   F & 1 & J 
  \end{Bmatrix}} 
   d_e \mathcal{E}\langle \gamma J||\Sigma_0^{\text{EDM},1} || \gamma J \rangle \tag{\text{A11}}
\end{align*}
where the reduced matrix element $\langle \gamma J||\Sigma_0^{\text{EDM},1} || \gamma J \rangle$ is the linear Stark matrix element calculated neglecting the hyperfine structure. The dependence on $M_F$ is contained in the Clebsh-Gordan coefficient $C_{FM_F10}^{FM_F}$. This coefficient equals to \cite{Var88}
\begin{equation*}
\label{A12}
C_{FM_F10}^{FM_F}=\frac{M_F}{[F(F+1)]^{1/2}}.  \tag{\text{A12}}
\end{equation*}

\end{document}